\documentclass[proceedings]{JHEP}
\PrHEP{PrHEP hep2001}                   
\conference{International Europhysics Conference on HEP}
\usepackage{epsfig}                   

\title{Non-adiabatic oscillations of supernova neutrinos}

\author{M. Kachelrie{\ss}\\                
        TH Division, CERN, CH-1211 Geneva 23\\ 
        E-mail: \email{Michael.Kachelriess@cern.ch}}

\abstract{
The level-crossing probability, local and global adiabaticity conditions 
are discussed for 2-flavour neutrino oscillations in matter with
arbitrary mixing  
angles $\vartheta$. Different approximations for the survival probability of
supernova neutrinos are compared. Results of a combined likelihood
analysis of the observed SN~1987A neutrino signal and of the latest solar
neutrino data including the recent SNO CC measurement are presented.}

\def\i{{\rm i}}
\def\d{{\rm d}}
\def\e{{\rm e}}

\def\ap{\approx}
\def\F{{\cal F}}
\def\G{{\cal G}}

\def\PLZ{P_{\rm LSZ}}
\def\Ee{\langle E_{\bar\nu_{\rm e}}\rangle}
\def\Eh{\langle E_{\bar\nu_{\rm h}}\rangle}
\def\Eb{E_{\rm b}}

\def\t{\vartheta}
\def\tm{\vartheta_{\rm m}}
\def\Dm{\Delta_{\rm m}}

\newcommand{\lsim}   {\mathrel{\mathop{\kern 0pt \rlap
  {\raise.2ex\hbox{$<$}}}
  \lower.9ex\hbox{\kern-.190em $\sim$}}}
\newcommand{\gsim}   {\mathrel{\mathop{\kern 0pt \rlap
  {\raise.2ex\hbox{$>$}}}
  \lower.9ex\hbox{\kern-.190em $\sim$}}}

\def\be{\begin{equation}}
\def\ee{\end{equation}}
\def\ba{\begin{eqnarray}}
\def\ea{\end{eqnarray}}

\begin{document}

\section{Neutrino evolution: resonance  and adiabaticity conditions,
  maximal violation of adiabaticity}
We consider neutrino oscillations in a two flavour scenario and
label the heavier neutrino mass eigenstate with ``2''. Then
$\Delta=m_2^2-m_1^2$ is positive and the vacuum mixing angle $\t$
is in the range $[0\!:\!\pi/2]$. As starting point for our discussion,
we use the evolution equation for the medium states $\tilde\psi$~\cite{Fr00} 
\be  \label{S}
 \frac{\d}{\d\tm}
 \left( \begin{array}{c} \tilde\psi_1 \\ \tilde\psi_2 \end{array}
 \right) =
 \left( \begin{array}{cc} {\i}\frac{\Dm}{4E\tm^\prime} & -1 \\
                          1 & -{\i}\frac{\Dm}{4E\tm^\prime}  \end{array}
 \right)
 \left( \begin{array}{c} \tilde\psi_1 \\ \tilde\psi_2 \end{array}
 \right) \,,
\ee
where $\Dm = \{\left(A-\Delta\cos 2\t\right)^2 + (\Delta\sin 2\t)^2 \}^{1/2}$
denotes the difference between the effective masses of the two
(active) neutrino states in matter, $E$ is their energy and
$A=2EV=2\sqrt{2}G_F N_e E$ is the induced mass squared for the electron neutrino. Furthermore, $\tm$ is the mixing angle in matter and 
$\tm^\prime=\d\tm/\d r$. Since anti-neutrinos feel a potential $V$ with
the opposite sign than neutrinos, formulae derived below
for neutrinos become valid for anti-neutrinos after the substitution
$\t\to\pi/2-\t$.

The traditional condition for an adiabatic evolution of a neutrino
state along a certain trajectory is that the diagonal entries of the
Hamiltonian in Eq.~(\ref{S}) are large with respect to the
non-diagonal ones, $|\Dm|\gg |4E\tm^\prime|$. This condition measures
indeed how strong adiabaticity is {\em locally\/} violated. Therefore, the
point of maximal violation of adiabaticity (PMVA)
 is given by the minimum of $\Dm/\tm^\prime$. In the following, we
will concentrate on power-law like  potential profiles, $A\propto r^n$.
This type of profile does not only contain the case $n \ap -3$ typical
for supernova envelopes, but also the exponential profile of the sun
in the limit $n\to\pm\infty$. Moreover, it allows discussing which
features of neutrino oscillations are generic and which ones are
specific for the linear profile $n=1$ usually discussed. 
For $A(r)\propto r^n$, the minimum of $\Dm/\tm^\prime$ is at
\ba  \label{eq_pmva}
 & & \cot(2\tm-2\t) + 2\cot(2\tm) \\ && 
 -\frac{1}{n} \, \left[ \cot(2\tm-2\t) - \cot(2\tm) \right] = 0 \,.
\nonumber
\ea
For $n=1$, the PMVA is indeed at $\tm=\pi/4$ for all $\t$. Thus, in
the region where the resonance point $\tm=\pi/4$ is well-defined, they
coincide. In the general case, $n\neq 1$, the PMVA agrees however only for
$\t=0$ with the resonance point.

\begin{figure}
\vskip-0.5cm
\epsfig{file=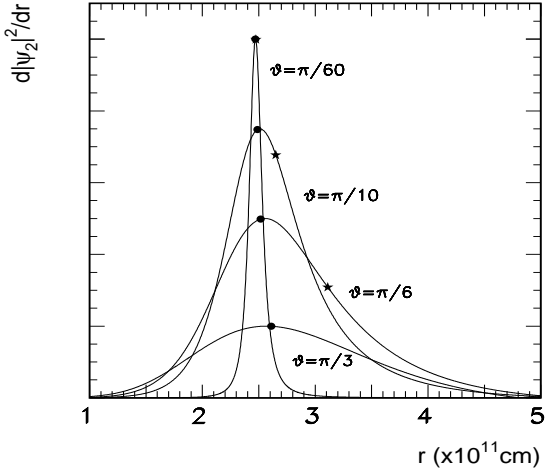,height=7.cm,width=8.cm,angle=0}
\caption{\label{pmva2}
  Change of the survival probability $\d p(r)/\d r$ of a neutrino
  produced at $r=0$ as $\tilde\nu_2$ together with the point of maximal
  violation of adiabaticity (dot) and the resonance point (star) for a
  power law profile $A\propto r^{-3}$. The height of the different
  curves is rescaled.}
\end{figure}

In Fig.~\ref{pmva2}, we show the the change of the survival probability, 
$\d p(r)/\d r=\d|\tilde\psi_2(r)|^2/\d r$ for a neutrino produced at $r=0$ as
$\tilde\nu_2$, together with the PMVA predicted by Eq.~(\ref{eq_pmva})
and the resonance point for a power law profile $A\propto r^{-3}$.  The
resonance condition predicts a transition in lower-density regions
than the PMVA, until for 
$\t=\pi/4$ the resonance point reaches $r=\infty$ and the concept of a
resonant transition breaks down completely. Moreover, the 
crossing probability becomes less and less localized
near the PMVA for larger mixing angles $\t$.

\begin{figure}
\vskip-0.5cm
\epsfig{file=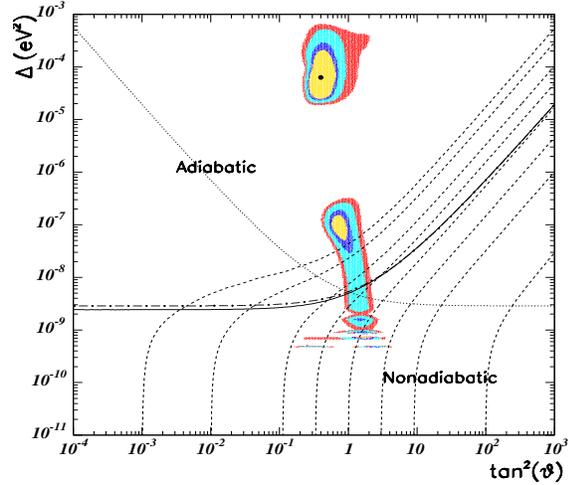,height=7.cm,width=8.cm,angle=0}
\caption{\label{border2}
  Contours of constant anti-neutrino survival probability (dashed)
  together with the borderline Eq.~(\ref{b})  between adiabatic and
  non-adiabatic regions using Eq.~(\ref{i}) with $n=-3$ (solid) and
  $n\to\pm\infty$ (dash-dotted) in $f(\t)$ for the SN profile given in the
  text; the dotted line shows the borderline for neutrinos.}
\end{figure}

Let us now discuss the condition for the adiabatic evolution of a
neutrino state along a trajectory from the core of a star to the vacuum. 
While the condition $|\Dm|\gg |4E\tm^\prime|$ indicates whether 
adiabaticity is locally violated, we need now a {\em global\/} criterion 
that measures the cumulative non-adiabatic effects along the trajectory
from $\tm\ap\pi/2$ to $\t$. For a non-adiabatic evolution of the
neutrino state we require that
\be
 \bigg| \!\int_{\pi/2}^\t\d\tm \: \tilde\psi_1  \bigg| = \varepsilon\:
  \bigg| \!\int_{\pi/2}^\t\d\tm \: \frac{4E\tm'}{\Dm} \: \tilde\psi_2
\bigg|
\ee
with $\varepsilon\ll 1$. Then the border between the adiabatic and
non-adiabatic regions is given by
\be
\label{b}
\frac{\Delta}{E} = \left\{ \varepsilon\:
\frac{f(\t)}{\sin^2(2\t)(1-\sin\t)} \:
                  \frac{2n (2V_0)^{1/n}}{R_0}\right\}^{\frac{n}{n+1}}
\,,
\ee
with
\ba  \label{i}
 f(\t) &=&  |\int_{\pi/2}^\t\d\tm \sin\tm \left[ \sin(2\tm) \right]^{2+1/n}
\nonumber\\ & & \phantom{|\int_{\pi/2}^\t}
                \times \left[ \sin(2\tm-2\t) \right]^{1-1/n} |  \,.
\ea
Fig.~\ref{border2} shows the excellent agreement between our prediction
for the border between the non-adiabatic and adiabatic regions for 
anti-neutrinos with energy $E=20$~MeV and a profile typical for
supernova envelopes, 
$V(r)= 1.5\times 10^{-9}$~eV $(10^9~{\rm cm}/r)^{3}$, and the one
following from  the contours of constant survival
probability $P_{ee}$ (dashed lines) of the neutrino eigenstate $\tilde\nu_2$
obtained by solving the Schr{\"o}dinger equation~(\ref{S}). A
comparison of the solid ($n=-3$) and the dash-dotted line
($n\to\pm\infty$) shows moreover that $f(\t)$ depends only weakly on $n$.

\section{The crossing probability in the WKB formalism}
The leading term to the crossing probability $\PLZ$ within the WKB
formalism is in the  ultra-relativistic limit and omitting an overall
phase given by
\be  \label{start}
 \ln\PLZ = -\frac{1}{E}\: \Im\int_{x_1(A_1)}^{x_2(A_2)} \d x \:\Dm \,,
\ee
where $A_2=\pm\Delta e^{2{\i}\t}$ are the branch points of
$\Dm$ in the complex $x$ plane and and $x_2$ can be chosen
arbitrarily either on the positive or negative real $x$ axis.
The usual choice, $A_1=\Delta C$, allows to
express $\ln\PLZ$ as the product of the adiabaticity parameter
$\gamma$ evaluated at the resonance point and a correction function
$\F_n$~\cite{Ku89} that can be represented as a hypergeometric
function $_2F_1$~\cite{Ka01b}.

Another representation for the crossing probability which is valid for
all $\t$ uses as integration path in the complex $x$ plane the part of
a circle of radius $\Delta$ centred at zero and starting from
$A_1=\Delta$ and ending at $A_2=\Delta\e^{2{\i}\t}$. For $A=A_0
(r/R_0)^n$, one can factor out
the $\t$ dependence of $\PLZ$ into functions $\G_n$~\cite{Ka01b},
\be
 \ln\PLZ = -\kappa_n \G_n(\t) \,,
\ee
where
\be
 \kappa_n =
                 \left( \frac{\Delta}{E} \right) \:
                 \left( \frac{\Delta}{A_0} \right)^{1/n} R_0
\ee
is independent of $\t$ and
\be
 \G_n(\t) = \left| \; \Re
 \int_0^{2\t/n} \d\varphi \:\e^{{\i}\varphi}
 \left[\left(\e^{{\i} n\varphi}-C\right)^2 + S \right]^{1/2}  \right| 
\ee
with $C=\cos 2\t$ and $S=\sin 2\t$.

\section{Neutrino oscillations in supernova envelopes}
In the analysis of neutrino oscillations, the potential profile $A(r)$
of supernova (SN) envelopes is often approximated by a power law with
$n\ap -3$, and $V(r)= 1.5\times 10^{-9}$~eV $(10^9~{\rm cm}/r)^{3}$.
A comparison of the results of a numerical solution of
the Schr{\"o}dinger equation~(\ref{S}) with the analytical calculation
of $P_{\bar e \bar e}$ using the $\G_{-3}$ functions shows very good
agreement for this profile; only tiny deviations in the region
$\Delta/(2E)\sim 10^{-17}$~eV have been found in \cite{Ka01c}. By
contrast, all other approximations used hitherto in  the
literature fail in some part of the $\tan^2\t$--$\Delta$ plane:  
while the use of $\F_1=1$ together with $A\propto r^{-3}$ describes
correctly the crossing
probability for small mixing in the resonant region, the errors become
larger for larger mixing until this approximation fails completely in
the non-resonant region.  The correction function $\F_\infty$ used for
$n=-3$ describes quite accurately the most interesting region of large
mixing as well as the non-resonant region, but does not reproduce the
correct shape of the MSW triangle.  

\begin{figure}
\vskip-0.5cm
\epsfig{file=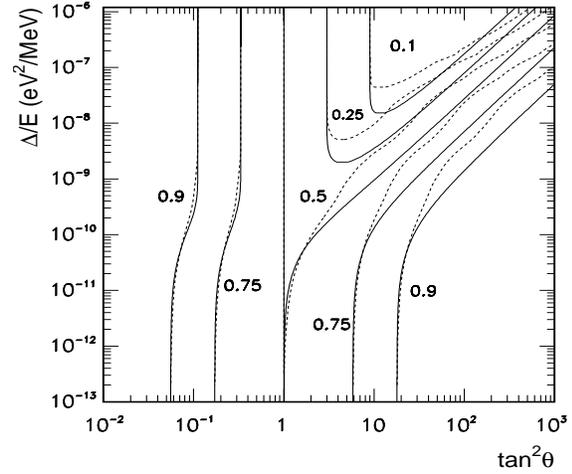,height=7.cm,width=8.cm,angle=0}
\caption{\label{woosley}
Comparison of the contours of constant survival probability $P_{\bar
    e\bar e}$ calculated numerically for a $M=20M_\odot$ progenitor star
  (dotted lines) and calculated for $A\propto r^{-3}$
  with the $\PLZ$ approximation (solid lines). }
\end{figure}

More important is however to check how strong deviations of the true
SN progenitor profile $V(r)$ from a power-law profile may affect the
analytical results. Realistic progenitor profiles differ in two
aspects from a simple $1/r^3$ behaviour. First, the outer part
of the envelope has an onion like structure, and its chemical
composition, $Y_e(r)$, and thus also $V(r)$ changes rather sharply at
the boundaries of the various shells. Second, the density drops faster
in the outermost part of the envelope, becoming closer to an
exponential decrease. We calculated numerically $P_{\bar e\bar e}$ 
using profiles for different progenitor masses and stellar evolution
models. We found that $P_{\bar e\bar e}$ is well approximated
only in the non-resonant part by our analytical results for
the $1/r^3$ profile, independently of the details of the progenitor
profile, while $P_{\bar e\bar e}$ depends strongly on the
details of the progenitor profile in the resonant region. As an example, we
compare in Fig.~\ref{woosley} the $P_{\bar e\bar e}$ calculated
numerically for a $20M_\odot$ profile with the analytical results for our
standard SN profile.
Therefore a numerical solution of the
Schr{\"o}dinger equation~(\ref{S}) should be performed in the resonant
region, using a realistic profile for the particular progenitor star
considered.  However, a $1/r^3$ profile together with the WKB crossing 
probability is sufficient for the analysis of {\em anti-}neutrino
oscillations in the phenomenologically most interesting region
$\tan^2\t\lsim 5$ independent of the details of the progenitor envelope.

Next we present the results of a combined statistical analysis of the
neutrino signal of SN~1987A and of the complete set of solar neutrino
experiments~\cite{Ka01c}. Since the two data sets are statistically
independent and functions of the same two fit parameters, they can be
trivially combined,
\be
 \chi^2_{\rm tot}(\t,\Delta) =
 \chi^2_\odot(\t,\Delta)+\chi^2_{\rm SN}(\t,\Delta) \,.
\ee
Contours of constant confidence level (C.L.) are defined relative to the
minimum of $\chi^2_{\rm tot}$, where $\chi^2_\odot(\t,\Delta)$ was
calculated in Ref.~\cite{solar} for the solar data and $\chi^2_{\rm SN}
=-2\, {\cal L}(\t,\Delta)$ in Ref.~\cite{Ka01a} for the SN~1987A data.  
We consider the astrophysical parameters as known and minimize only
the two parameters $\t$ and $\Delta$.

\begin{figure}
\vskip-0.5cm 
\epsfig{file=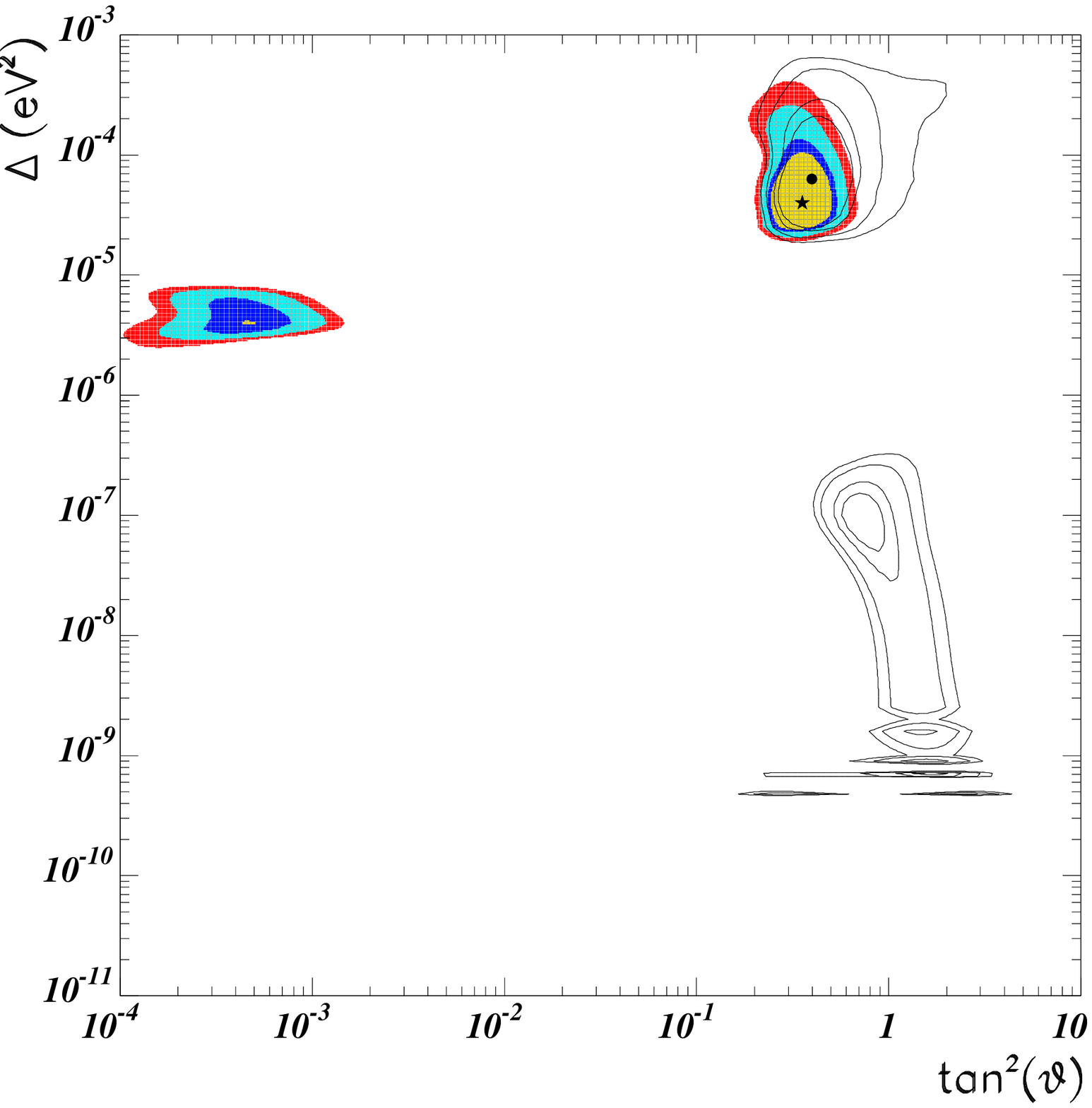,height=7.cm,width=8.cm,angle=0}
\vskip-0.5cm
\epsfig{file=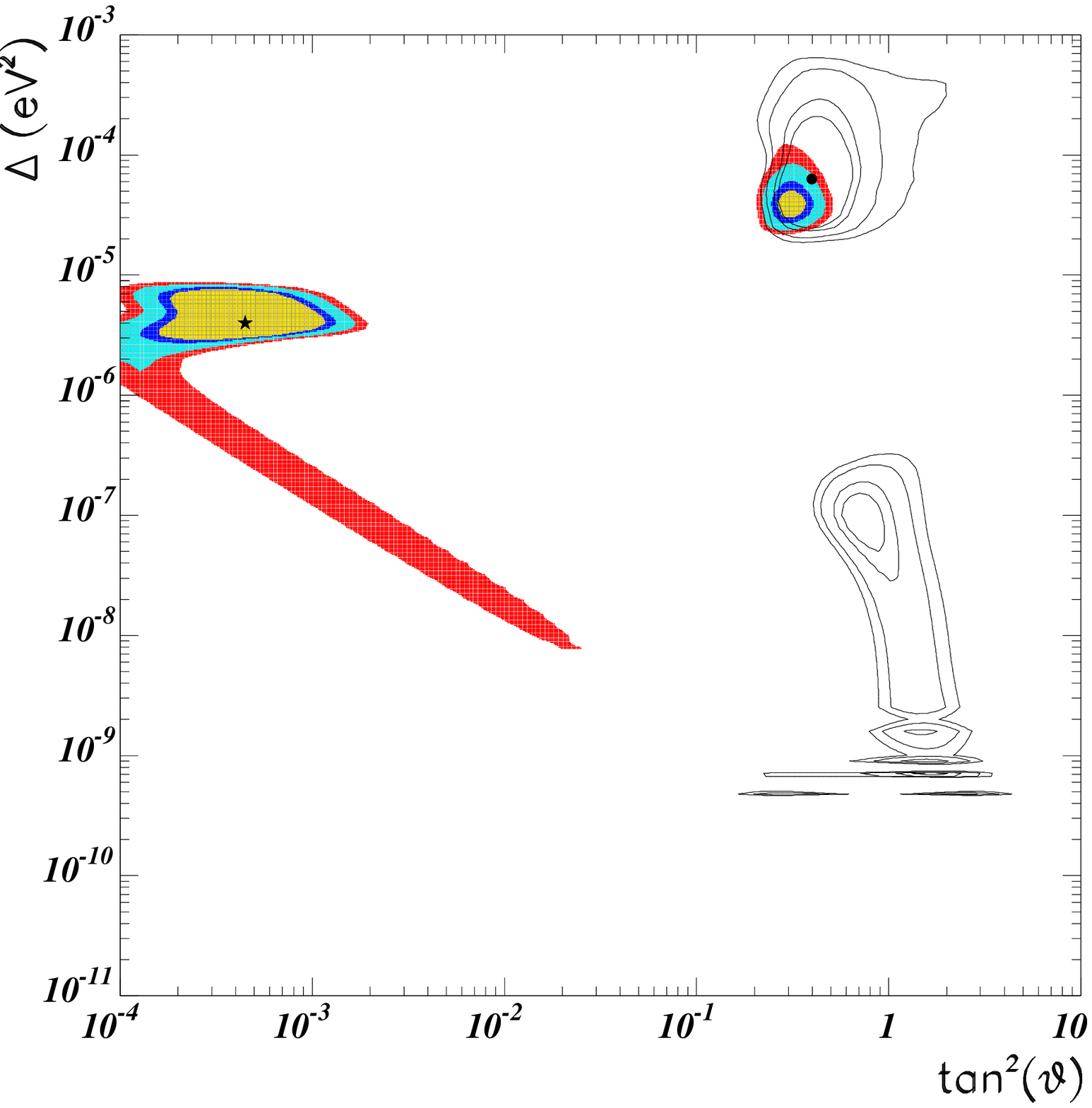,height=7.cm,width=8.cm,angle=0}
\caption{\label{3_14_tot}
  The 90, 95, 99 and 99.73\% C.L. contours of the combined fit of solar
  and SN~1987A data (coloured/grey) together with the contours of the solar
  data alone (solid lines);
  for $\tau=\Eh/\Ee=1.4$ (top) and $\tau=1.7$ (bottom).
  All figures for $\Eb=3\times 10^{53}$~erg and $\Ee=14$~MeV.}
\end{figure}

In Fig.~\ref{3_14_tot} we show the C.L. contours of the combined fit
for a rather representative set of astrophysical parameters, namely
binding energy $\Eb=3\times 10^{53}$~erg and $\Ee=14$~MeV. In this
case, the impact of the SN~1987A data on the standard solutions to the
solar neutrino problem is rather dramatic: the LOW-QVAC and VAC
solutions disappear for both assumed $\tau=\Eh/\Ee$ values; they are
excluded at more than 99.98\% even for $\tau=1.4$.
Moreover the size of the LMA--MSW solution decreases with increasing
$\tau$. The part of the LMA--MSW solution which is most stable against
the addition of the supernova data corresponds to the lowest $\Delta$
and
$\tan^2\t$ values, since these are favoured by Earth matter
regeneration effects. On the other hand the SMA--MSW region re-appears
extending, for increasing $\tau$, as a funnel towards the VAC solution
along the hypotenuse of the solar MSW triangle.  The combined best-fit
point (star) moves from the LMA--MSW region for $\tau=1.4$ to the SMA--MSW
solution for $\tau=1.7$.

\begin{figure}
\vskip-0.5cm
\epsfig{file=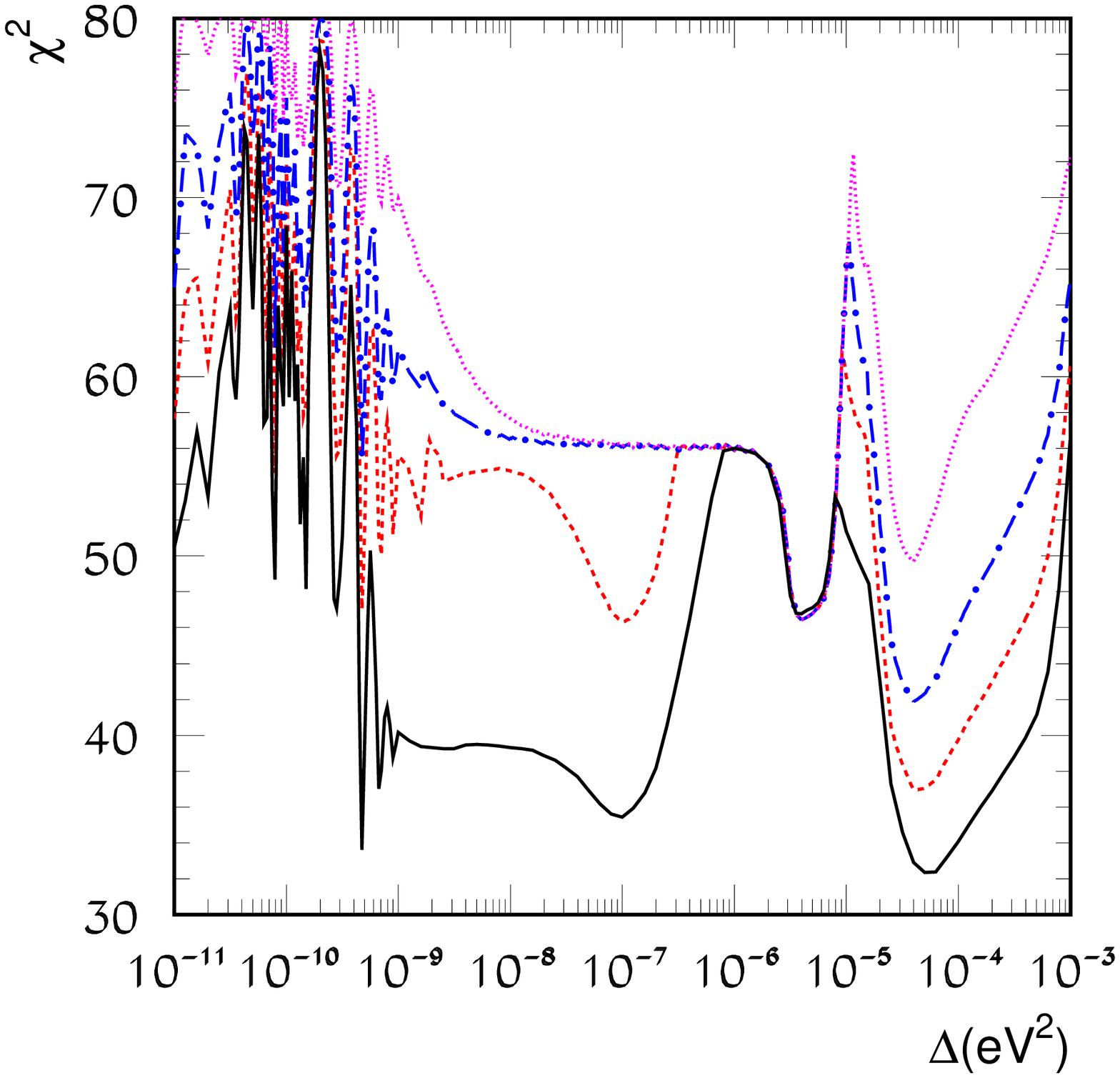,height=7.cm,width=8.cm,angle=0}
\vskip-0.5cm
\epsfig{file=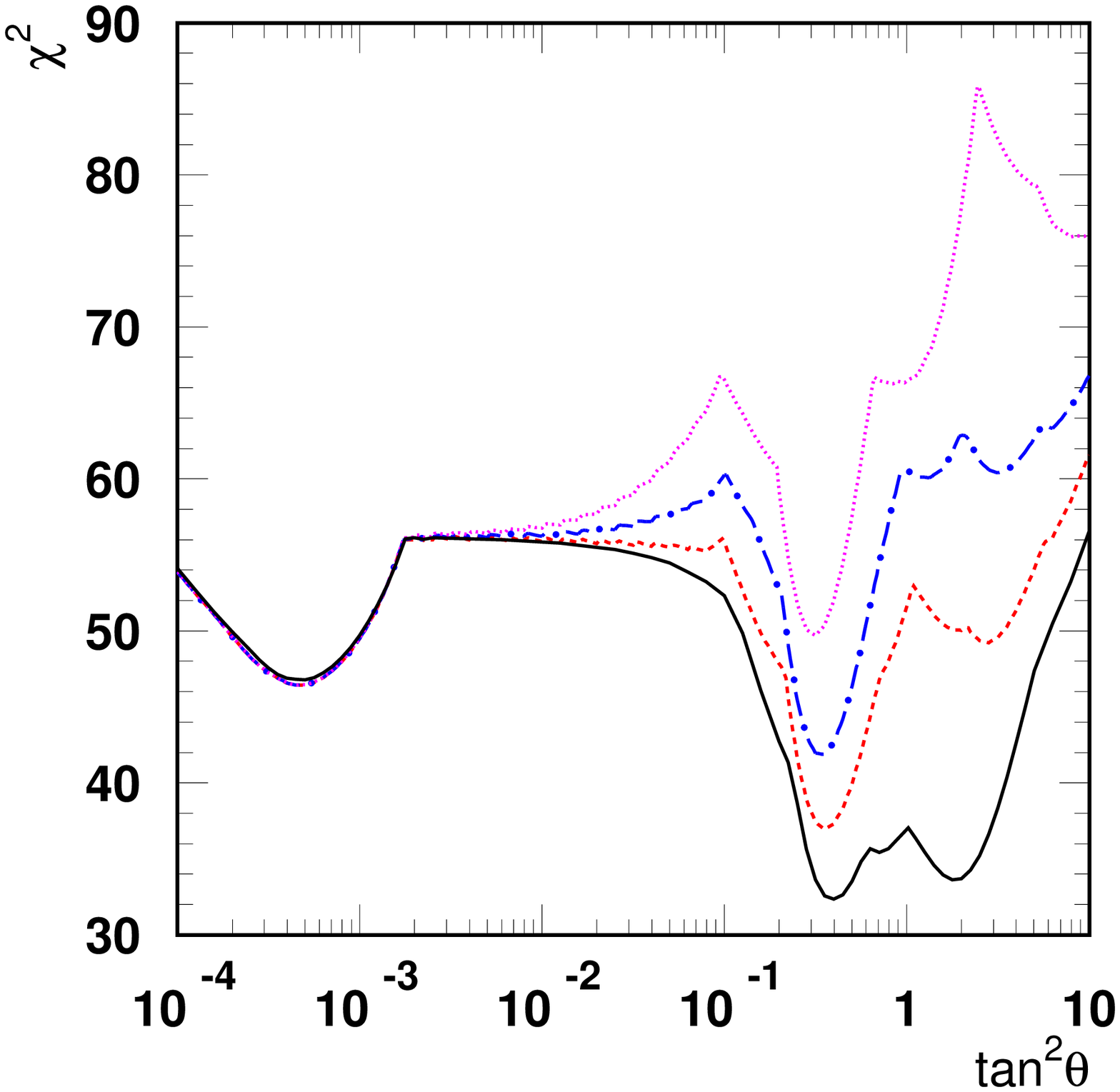,height=7.cm,width=8.cm,angle=0}
  \caption{\label{combdm}
    The solid curve indicates the $\chi_\odot^2$ of the solar neutrino
    data. The non-solid
    curves illustrate the effect of adding the SN~1987A data, which
    worsens the status of large mixing-type solutions; marginalized
    with respect to $\t$ (top) and to $\Delta$ (bottom), respectively.}
\end{figure}

In Fig.~\ref{combdm} we illustrate in a global way the relative status
of various oscillation after adding the SN~1987A data. For each
$\Delta$ value we have optimized the $\chi^2$ with respect to $\t$ in
the top and with respect to $\Delta$ in the bottom panel. 
The solid curve indicates $\chi_\odot^2$, the non-solid
curves correspond to the case where the SN~1987A data are included.
The dash-dotted line is for $\Eb=3\times 10^{53}$~erg,
$\tau=1.4$ and $\Ee=14$~MeV.  The dashed line is for
$\Eb=3\times 10^{53}$~erg, $\tau=1.4$
and $\Ee=12$~MeV.  The dotted line is for $\Eb=3\times 10^{53}$~erg,
$\tau=1.7$ and $\Ee=14$~MeV. Here we have adjusted an arbitrary constant
which appears when combining the minimum likelihood-type SN~1987A
analysis with the solar $\chi^2$ data analysis in such a way that the
SMA
solution gets unaffected by the SN~1987A data. One notices that the
effect of adding SN~1987A data is always to worsen the status of
the large-mixing angle solutions. Within each such
curve one can compare the relative goodness of various solutions,
however different curves should not be qualitatively compared.

\section{Summary and Discussion}
We have discussed non-adiabatic neutrino oscillations in general
power-law potentials $A\propto x^n$. We found that the resonance point
coincides only for a linear profile with the point of maximal
violation of adiabaticity. We presented the correct boundary between
the adiabatic and non-adiabatic regime for all $\t$ and $n$ as well as
a new method to calculate the crossing probability also in the non-resonant
regime.

Performing a combined likelihood analysis of the observed neutrino
signal of SN~1987A and solar neutrino data including the SNO CC
measurement, we found that the supernova data offer additional
discrimination power between the different solutions
of the solar neutrino puzzle.
Unless all relevant supernova parameters lie close
to their extreme values found in simulations, the status of the LMA
solutions deteriorates, although the LMA--MSW solution may still
survive as the best combined fit  for acceptable
choices of astrophysical parameters. In particular, SN~1987A data generally
favour its part with smaller values of $\t$ and $\Delta$.
In contrast the vacuum or ``just-so'' solution is excluded and the LOW
solution is significantly disfavoured
for most reasonable choices of astrophysics parameters.
The SMA--MSW solution is absent at
about the $3\sigma$-level if solar data only are included but may
reappear once
SN~1987A data are added,  due to the worsening of the LMA  type solutions.

Finally, one should not forget that in the solar case, a
well-tested standard solar model exists whose errors 
are accounted for in the fit. In contrast
there is no ``standard model'' for type II supernovae and
therefore also no well-established average values and error estimates
for the relevant astrophysical parameters.

\section*{Acknowledgements}
\vskip-0.19cm
I am grateful to A.~Strumia, R. Tom{\`a}s and J.W.F. Valle for
fruitful collaborations on which this talk is based on, to
C. Pe{\~n}a-Garay for sharing insight and data, and to
M.C.~Gonzalez-Garcia for a lively discussion at the EPS conference.


\begin{thebibliography}{99}

\bibitem{Fr00}
A.~Friedland,
\prd{64}{2001}{013008}.


\bibitem{Ku89}
T.K.~Kuo and J.~Pantaleone,
\prd{39}{1989}{1930}.

\bibitem{Ka01b}
M.~Kachelrie{\ss} and R.~Tom\`as,
\prd{64}{2001}{073002}.

\bibitem{Ka01c}
M.~Kachelrie{\ss}, A.~Strumia, R.~Tom\`as and J.W.F.~Valle,
\hepph{0108100}.

\bibitem{solar} 
We used the data set corresponding to Fig.~6b
of P.~Creminelli, G.~Signorelli and A.~Strumia,
 \hepph{0102234}v2.

\bibitem{Ka01a}
M.~Kachelrie{\ss}, R.~Tom\`as and J.W.F.~Valle,
\jhep{01}{2001}{030}.

\end{thebibliography}
\end{document}